\documentclass[journal]{IEEEtran}

\usepackage{color}
\usepackage[cmex10]{amsmath}
\usepackage{amssymb}
\usepackage{dsfont}
\usepackage{epsfig}
\usepackage{stfloats}
\usepackage{cite}
\usepackage{subfigure}

\newtheorem{lemma}{\it Lemma}
\newtheorem{theorem}{\it Theorem}
\newtheorem{remark}{\it Remark}
\newtheorem{corollary}{\it Corollary}
\begin{document}

\title{\LARGE Secure Two-Way Communications via Intelligent Reflecting Surfaces}

\author{Lu Lv,~\IEEEmembership{Member,~IEEE}, Qingqing Wu,~\IEEEmembership{Member,~IEEE}, Zan Li,~\IEEEmembership{Senior~Member,~IEEE},\\ Naofal Al-Dhahir,~\IEEEmembership{Fellow,~IEEE}, and Jian Chen,~\IEEEmembership{Member,~IEEE}\vspace{-5.8mm}
\thanks{This work was partially supported by the National Natural Science Foundation of China under Grants 61901313, 61771366, 61825104, and 61941105, the Open Research Fund of National Mobile Communications Research Laboratory, Southeast University, under Grant 2020D07, the Natural Science Basic Research Plan of Shaanxi Province under Grant 2020JQ-306, and the the China Postdoctoral Science Foundation under Grants BX20190264 and 2019M650258. The editor coordinating the review of this paper and approving it for publication was N. Miridakis. {\it (Corresponding author: Jian Chen.)}}
\thanks{Lu Lv is with the State Key Laboratory of Integrated Services Networks, Xidian University, Xi'an 710071, China, and also with the National Mobile Communications Research Laboratory, Southeast University, Nanjing 210096, China (e-mail: lulv@xidian.edu.cn).}
\thanks{Qingqing Wu is with the State Key Laboratory of Internet of Things for Smart City, University of Macau, Macau 999078, China (e-mail: qingqingwu@um.edu.mo).}
\thanks{Zan Li and Jian Chen are with the State Key Laboratory of Integrated Services Networks, Xidian University, Xi'an 710071, China (e-mail: zanli@xidian.edu.cn; jianchen@mail.xidian.edu.cn).}
\thanks{Naofal Al-Dhahir is with the Department of Electrical and Computer Engineering, The University of Texas at Dallas, Richardson, TX 75080, USA (e-mail: aldhahir@utdallas.edu).}
}
\markboth{IEEE Communications Letters}{}
\maketitle

\begin{abstract}
  In this letter, we propose to exploit an intelligent reflecting surface (IRS) to secure multiuser two-way communications. Specifically, two end users simultaneously transmit their signals, where the signal of one user is exploited as information jamming to disrupt the reception of the other user's signal at an eavesdropper. A simple user scheduling scheme is developed to improve the secrecy rate. Analytical results in terms of a lower bound on the average secrecy rate and its scaling laws are derived to evaluate the secrecy performance and obtain valuable design insights. Simulation results are provided to confirm the accuracy of the theoretical analysis and validate the secrecy improvement of the proposed scheme over other baseline schemes.
\end{abstract}
\IEEEpeerreviewmaketitle

\section{Introduction}

Intelligent reflecting surface (IRS) is a promising technology to achieve energy and spectrally efficient communications \cite{QingqingWu_CM2020}. Typically, IRS is a planar surface that consists of a large number of passive and low-cost reconfigurable reflecting elements. Each element can be managed to adjust its phase and/or amplitude for the incident signal independently. This creates a smart and controllable signal propagation environment for improving the spectrum utilization \cite{Saman_TCOM2020,YuZhang_CL2020}, radio coverage, and broadband connectivity \cite{ZDing_WCL2020,TianweiHou_JSAC2020} for 6G wireless networks.

Thanks to its capability of reconfiguring wireless channels, IRS has great potential to enhance physical-layer security for wireless communications. By intelligently designing the phase shifts of the IRS, signals reflected by each reflecting element can be added constructively at the legitimate receiver to enhance its reception quality, while being received destructively at eavesdropper (Eve) to degrade its signal reception. Hence, an improved secrecy rate can be achieved. In \cite{XinrongGuan_WCL2020}, a secrecy rate maximization problem for an IRS assisted multiple-input single-output (MISO) system was investigated, where artificial noise is used to jam the Eve. In \cite{LimengDong_WCL2020}, an efficient alternating optimization algorithm was proposed to maximize the secrecy rate of multiple-input multiple-output transmission using IRS. In \cite{ZhengChu_WCL2020}, the transmit power subject to a secrecy rate constraint was minimized for an IRS based MISO downlink system. Assuming that Eve's channel is unknown, a joint transmit beamforming and cooperative jamming strategy for IRS enhanced secrecy transmission was developed in \cite{HuimingWang_SPL2020}.

All the existing literature on IRS-aided physical-layer security considered one-way communications. However, to our best knowledge, the potential of using IRS to secure  two-way communications has not been explored yet. Theoretically, the superposition of two-way communications via the IRS can result in an overlapping of multiple signals at Eve, which provides a new design opportunity to improve the physical-layer security by reusing both information signals of the user's direct path and the IRS reflecting path as jamming noise to degrade Eve's reception performance. The research on this topic is still in its infancy, and how to design secure IRS assisted two-way communications remains unknown. Motivated by the above considerations, as the first work, we investigate the exploitation of the IRS to secure two-way communications and reveal its achievable fundamental performance limits. The main contributions can be summarized as follows.
\begin{itemize}
  \item {\it Novel framework:} We propose a new secure IRS assisted two-way communication scheme to exploit information jamming for improving the secrecy rate. Specifically, two end users simultaneously transmit their signals via the IRS, where the signal of one user is utilized as a source of helpful information jamming to degrade the capability of Eve. Each user can remove the interfering signal (a copy that is previously transmitted by itself) via the self-interference cancellation technique without affecting its desired signal reception. Furthermore, a user scheduling scheme is also devised to exploit the multiuser diversity gain to enhance the secrecy rate.
  \item {\it Tractable analysis:} We propose an accurate approximation for deriving a closed-form lower bound on the average secrecy rate (ASR). To provide more useful insights on the impact of system parameters, we then derive ASR scaling laws with sufficiently large transmit power $P$, number of IRS reflecting elements $K$, and number of user pairs $N$.
  \item {\it Valuable insights:} Both analytical and numerical results reveal that: 1) The proposed secure IRS assisted two-way communication scheme can guarantee perfect transmission security and significantly outperform other baseline schemes in general. 2) The ASR for each signal increases at the rates of $\log P$, $2\log K$, and $\log\log N$ at sufficiently large $P$, $K$, and $N$. These results are highly promising since they indicate that the proposed scheme yields the same scaling laws as the non-secrecy transmission.
\end{itemize}

\section{System Model}

Consider an IRS assisted two-way communications system, which consists of $N$ pairs of end users, denoted by $\{\text{A}_n,\text{B}_n\}$ for $n=1,\dots,N$, an IRS, and a passive Eve. We assume no direct links between $\text{A}_n$ and $\text{B}_n$, due to blockages, which motivates the deployment of an IRS to establish communication links. Eve is assumed to be located close to the users and IRS, so that it can overhear the confidential information from both the direct and reflecting channels. The IRS has $K$ low-cost reconfigurable reflecting elements, and each element can reflect a phase shifted version of the incident signal independently, to enhance the signal reception quality at all users while reducing the information leakage to Eve.

All the channels experience quasi-static block fading. In a fading block, the channel vectors between $\text{A}_n$/$\text{B}_n$ and IRS are denoted by $\mathbf{h}_n$ and $\mathbf{g}_n$, respectively, the channel vector between the IRS and Eve is denoted by $\mathbf{h}_e$, and the fading gains between $\text{A}_n$/$\text{B}_n$ and Eve are denoted by $h_{ne}$ and $g_{ne}$, respectively. We assume channel reciprocity for user-to-IRS and IRS-to-user channels. Each entry of $\mathbf{h}_n$, $\mathbf{g}_n$, and $\mathbf{h}_e$, as well as $h_{ne}$ and $g_{ne}$ are assumed to be independent complex Gaussian distributed with zero mean and unit variance.

We assume that the users know the instantaneous CSI of $\mathbf{h}_n$ and $\mathbf{g}_n$, and the IRS knows the channel phase values of $\mathbf{h}_n$ and $\mathbf{g}_n$. However, the users only know the average CSI of $h_{ne}$ and $g_{ne}$, due to the passive nature of Eve (which means that Eve only listens but does not transmit, and thus, it is difficult for users to obtain the instantaneous CSI of the Eve).

The main goal of this letter is to exploit the IRS potentials (i.e., beamforming gain for end users and jamming capability for Eve) to enhance physical-layer security, quantify the fundamental performance limits, and provide useful guidelines for secure network design.

\subsection{Proposed Scheme}

As for the bidirectional information exchange, the two end users both work in a full-duplex mode to perform simultaneous transmission and reception. Assuming that $\text{A}_n$ and $\text{B}_n$ are scheduled. $\text{A}_n$ and $\text{B}_n$ transmit the unit-power signals $s_1$ and $s_2$, respectively, and the IRS reflects incident signals with a negligible delay. As indicated by \cite{Basar_Access2019,QingqingWu_TCOM2020}, the IRS cascaded channel is a multiplication of three terms, including transmitter-to-element $k$ channel, IRS reflection, and element $k$-to-receiver channel. The $K$-element IRS performs a linear mapping from the incident signal vector to a reflected signal vector by a $K\times K$ diagonal reflecting matrix $\Theta$. $\text{A}_n$ receives a superposition of the desired signal, the self-interference, and the residual loop-interference (RLI) via the IRS, given by
\begin{equation}
\label{FD-y1}
  y_{\text{A}_n}=\sqrt{\beta_\text{I-AB}}\mathbf{h}_n^T\Theta\mathbf{g}_ns_2 +\sqrt{\beta_\text{I-AA}}\mathbf{h}_n^T\Theta\mathbf{h}_ns_1 +l_{\text{A}_n}+n_{\text{A}_n}.
\end{equation}
In \eqref{FD-y1}, $\beta_\text{I-AB}=\frac{G_u^2S^2}{d_{\text{A}_n}^\alpha d_{\text{B}_n}^\alpha}$ is the pathloss from $\text{B}_n$ to $\text{A}_n$ via the $k$th surface element \cite{QingqingWu_TCOM2020},\footnote{The reflection amplitude is a function of the grazing angle theoretically. This effect, however, can be mitigated by proper element design in practice, e.g., using a thin substrate. It has been verified by experiments in \cite{Goldsmith_2005} that the effect of the grazing angle on the reflection amplitude is negligible for many cases with careful element design.} where $G_u$ is the antenna gain of the end users, $S$ is the size of each surface element, $d_{\text{A}_n}$ and $d_{\text{B}_n}$ are the distances from the IRS to $\text{A}_n$ and $\text{B}_n$ with $\alpha$ being the pathloss exponent. In addition, $\beta_\text{I-AA}=\frac{G_u^2S^2}{d_{\text{A}_n}^\alpha d_{\text{A}_n}^\alpha}$ is the pathloss from $\text{A}_n$ to itself via the IRS, $l_{\text{A}_n}$ is the RLI, and $n_{\text{A}_n}$ is the additive white Gaussian noise (AWGN) at $\text{A}_n$.

Since $\text{A}_n$ knows $\mathbf{h}_n$, $\Theta$, and $\beta_\text{I-AA}$, it can completely cancel its self-interference, i.e., the second term in \eqref{FD-y1}. After that, $\text{A}_n$ decodes its desired signal with received signal-to-interference-plus-noise ratio (SINR) given by $\gamma_{n1}=\frac{P\beta_\text{I-AB}|\mathbf{h}_n^T\Theta\mathbf{g}_n|^2}{ |l_{\text{A}_n}|^2+\sigma_0^2}$, where $P$ is the power of each signal and $\sigma_0^2$ is the variance of the AWGN. In this work, we adopt the RLI model according to \cite{Saman_TCOM2020}, where $l_{\text{A}_n}$ is a random variable with zero mean and variance of $\sigma_l^2$, which has similar characteristics to the AWGN. Similarly, the received SINR to decode $s_1$ at $\text{B}_n$ is given by $\gamma_{n2}=\frac{P\beta_\text{I-BA}|\mathbf{g}_n^T\Theta\mathbf{h}_n|^2}{ |l_{\text{B}_n}|^2+\sigma_0^2}$, where $\beta_\text{I-BA}=\frac{G_u^2S^2}{d_{\text{A}_n}^\alpha d_{\text{B}_n}^\alpha}$ and $l_{\text{B}_n}$ is the RLI with zero mean and variance of $\sigma_l^2$.

The received signal at the Eve can be expressed as
\begin{align}
\label{FD-ye}
  y_e&=\big(\sqrt{\beta_\text{I-Ae}}\mathbf{h}_e^T\Theta\mathbf{h}_n+ \sqrt{\beta_\text{D-Ae}}h_{ne}\big)s_1\nonumber\\
  &\quad+\big(\sqrt{\beta_\text{I-Be}}\mathbf{h}_e^T\Theta\mathbf{g}_n+ \sqrt{\beta_\text{D-Be}}g_{ne}\big)s_2+n_e.
\end{align}
In \eqref{FD-ye}, $\beta_\text{I-Ae}=\frac{G_uG_eS^2}{d_e^\alpha d_{\text{A}_n}^\alpha}$ is the pathloss from $\text{A}_n$ to the Eve via the $k$th surface element, where $G_e$ is the antenna gain of the Eve and $d_e$ is the distance from the IRS to the Eve. $\beta_\text{D-Ae}=\frac{G_uG_e}{d_{\text{A}_ne}^\alpha}$ is the pathloss from $\text{A}_n$ to the Eve, where $d_{\text{A}_ne}$ is the distance from $\text{A}_n$ to the Eve. $\beta_\text{I-Be}=\frac{G_uG_eS^2}{d_e^\alpha d_{\text{B}_n}^\alpha}$ is the pathloss from $\text{B}_n$ to the Eve via the $k$th surface element. $\beta_\text{D-Be}=\frac{G_uG_e}{d_{\text{B}_ne}^\alpha}$ is the pathloss from $\text{A}_n$ to the Eve. $n_e$ is the AWGN at the Eve.

Based on \eqref{FD-ye}, Eve tries to intercept both signals $s_1$ and $s_2$ using the successive interference cancellation technique \cite{Lu_TCOM2020}. Without loss of generality, we assume that Eve first decodes $s_1$ by treating $s_2$ as noise (while the other case can be treated similarly), yielding the received SINR as
\begin{equation}
\label{gamma-e-s1}
  \gamma_{e1}=\frac{P|\varphi_{ne}|^2}{P|\psi_{ne}|^2+\sigma_0^2},
\end{equation}
where $\varphi_{ne}=\sqrt{\beta_\text{I-Ae}} \mathbf{h}_e^T\Theta\mathbf{h}_n+\sqrt{\beta_\text{D-Ae}}h_{ne}$ and $\psi_{ne}=\sqrt{\beta_\text{I-Be}} \mathbf{h}_e^T\Theta\mathbf{g}_n+\sqrt{\beta_\text{D-Be}}g_{ne}$. If Eve can correctly decode $s_1$ (which means that $\gamma_{e1}\geq\gamma_{n1}$), then it can remove $s_1$ from its observations and decode $s_2$ in an interference-free manner. Otherwise, $s_1$ can be viewed as useful interference to decrease the received SINR of $s_2$ without any jamming signal injection. Thus, the received SINR to decode $s_2$ is obtained by
\begin{equation}
\label{gamma-e-s2}
  \gamma_{e2}=\begin{cases}
    \frac{P|\psi_{ne}|^2}{P|\varphi_{ne}|^2+\sigma_0^2}, & \text{if}\ \gamma_{e1}<\gamma_{n1}, \\
    \frac{P|\psi_{ne}|^2}{\sigma_0^2}, & \text{if}\ \gamma_{e1}\geq\gamma_{n1}.
  \end{cases}
\end{equation}

\subsection{User Scheduling}

Since the instantaneous CSI of Eve is absent, we propose a user scheduling scheme based on the instantaneous CSI of legitimate channels. To guarantee user scheduling fairness and improve the users' achievable rates, the user pair can be scheduled as $n^\ast=\arg\max_{n=1,\dots,N}\zeta_n$, where $\zeta_n=|\mathbf{h}_n^T\Theta\mathbf{g}_n|^2 =|\mathbf{g}_n^T\Theta\mathbf{h}_n|^2$. Thus, the probability of scheduling any user pair is $\frac1N$, since  $\Pr(n^\ast=n)=\Pr\big(\bigcap_{m\neq n}(\zeta_n>\zeta_m)\big)=\int_0^\infty\prod_{m\neq n}F_{\zeta_m}(x)f_{\zeta_n}(x)dx =\frac1N$.

Accordingly, the secrecy rate for $s_i$ is given by $R_{s_i}=\big\{\log(1+\gamma_{n^\ast i})-\log(1+\gamma_{ei})\big\}^+$, where $i\in\{1,2\}$.

\begin{remark}
  The proposed secure IRS assisted two-way communication scheme provide a new view on interference exploitation. The signal from one specific user is exploited as useful jamming to confuse Eve, as observed from \eqref{gamma-e-s1} and \eqref{gamma-e-s2}. This approach is cost-effective, since secrecy is guaranteed not by constructing intentional interference that consumes extra transmission power, but rather by reusing the signal that already exists in two-way communications.
\end{remark}

\begin{remark}
  Due to channel reciprocity, we know that $\gamma_{n1}$ and $\gamma_{n2}$ have the same expression, meaning that maximizing $\gamma_{n1}$ is equivalent to maximizing $\gamma_{n2}$. Thus, the user scheduling criterion can enhance the performance of both users.
\end{remark}

\begin{remark}
  With $R_{s_i}$, we observe that when $P$ becomes very large, $\gamma_{n^\ast1}$ goes to infinity while $\gamma_{e1}$ is upper bounded. Thus, we always have $\gamma_{n^\ast1}>\gamma_{e1}$, which indicates that perfect security for signal transmission of $s_1$ is achieved. Similarly, perfect security for signal transmission of $s_2$ is guaranteed.
\end{remark}

\subsection{Optimal Phase Shift Design}

To maximize the received SINRs at $\text{A}_n$ and $\text{B}_n$, the phase shifts at the IRS should be perfectly matched with the phases of $\mathbf{h}_n$ and $\mathbf{g}_n$, which yields $\zeta_n=\big(\sum_{k=1}^K|h_{nk}||g_{nk}|\big)^2$, where $|h_{nk}|$ and $|g_{nk}|$ are the amplitudes of the $k$th entries of $\mathbf{h}_n$ and $\mathbf{g}_n$, respectively. Generally, it is very challenging to derive an exact cumulative density function (CDF) of $\zeta_n$. To overcome this difficulty, we use the Gamma distribution to find a tight approximation for the CDF of $\zeta_n$, as characterized below.

\begin{lemma}
  The CDF of $\zeta_n$ can be approximated as $F_{\zeta_n}(x)\simeq\frac{1}{\Gamma(K\mu)}\gamma(K\mu,\frac{\sqrt{x}}{\nu})$, where $\mu=\frac{\pi^2}{16-\pi^2}$, $\nu=\frac{16-\pi^2}{4\pi}$, and $\gamma(\cdot,\cdot)$ and $\Gamma(\cdot)$ represent the lower incomplete Gamma function and the Gamma function.
\end{lemma}
\begin{IEEEproof}
  Using \cite[Lemma~1]{Saman_TCOM2020}, we know that the product $|h_{nk}||g_{nk}|$ can be approximated by a Gamma distribution with parameters $\mu$ and $\nu$. Then, a sum of $K$ independent and identically distributed Gamma random variables still follows a Gamma distribution with parameters $K\mu$ and $\nu$. After simple variable transformation, we prove the lemma.
\end{IEEEproof}

As shown in Section \ref{sec:simulation}, the approximation in Lemma~1 is accurate in the whole $P$ regime, which is more effective than the central limit theorem (CLT)-based approximation \cite{ZDing_WCL2020}.

\section{Performance Analysis}

To characterize the system performance and gain useful insights, we adopt the ASR as the performance metric, which has been widely adopted in the literature. In this section, we derive an accurate ASR lower bound and its scaling laws.

By definition, ASR is the statistical average of the secrecy rate over fading channels. Using Jensen's inequality (i.e., $\mathbb{E}[\max\{a,b\}]\geq\max\{\mathbb{E}[a],\mathbb{E}[b]\}$ since the Max function is convex and $a$, $b$ are random variables), the ASR for signal $s_i$ ($i=1,2$) can be lower bounded by $\bar{R}_{s_i}=\big\{\mathbb{E}[\log(1+\gamma_{n^\ast i})]-\mathbb{E}[\log(1+\gamma_{ei})]\big\}^+=\{Q_{\text{M},i}-Q_{\text{E},i}\}^+$.
The following theorem provides the lower bounds on the closed-form ASRs.

\begin{theorem}
  The ASR lower bounds for $s_1$ and $s_2$ achieved by the IRS assisted secure multiuser two-way communications scheme can be approximated by
  \begin{align}
    \bar{R}_{s_1}&\approx
    \bigg\{\sum_{n=1}^N\widetilde{\sum\limits_{m,M}}\frac{\rho_\text{AB}\sec^2x_m (1-F_{\zeta_{n^\ast1}}(\tan x_m))} {N(1+\rho_\text{AB}\tan x_m)}\nonumber\\
    & -\frac{\sigma_e^2}{\sigma_e^2-\sigma_{e'}^2}\bigg(e^{\frac{1}{\rho_0\sigma_{e'}^2}} \mathrm{Ei}\Big(\frac{-1}{\rho_0\sigma_{e'}^2}\Big)-e^{\frac{1}{\rho_0\sigma_e^2}} \mathrm{Ei}\Big(\frac{-1}{\rho_0\sigma_e^2}\Big)\bigg)\bigg\}^+, \\
    \bar{R}_{s_2}&\approx \bigg\{\sum_{n=1}^N\widetilde{\sum\limits_{m,M}}\frac{\rho_\text{BA}\sec^2x_m (1-F_{\zeta_{n^\ast1}}(\tan x_m))} {N(1+\rho_\text{BA}\tan x_m)}\nonumber\\
    &\quad -(J_1+J_2)\bigg\}^+,
  \end{align}
  where $\rho_\text{AB}=\frac{P\beta_\text{I-AB}}{\sigma_l^2+\sigma_0^2}$, $\rho_\text{BA}=\frac{P\beta_\text{I-BA}}{\sigma_l^2+\sigma_0^2}$, $\rho_0=\frac{P}{\sigma_0^2}$, $x_m=\frac{\pi}{4}(\theta_m+1)$, $\widetilde{\sum}_{m,M}=\sum_{m=1}^M\sqrt{1-\theta_m^2}$, $\theta_m=\cos(\frac{2m-1}{2M}\pi)$, $M$ is the the accuracy-complexity parameter of the Gauss-Chebyshev (G-C) quadrature, and $J_1$ and $J_2$ are given in Appendix.
\end{theorem}
\begin{IEEEproof}
  Please refer to Appendix.
\end{IEEEproof}

Theorem 1 provides an efficient method to evaluate the performance of the secure IRS assisted two-way communication scheme, since it consists of power functions, Gamma functions, trigonometric functions, which are easy to compute. However, these expressions do not yield more intuitive insights into the impacts of system parameters, such as the transmit power $P$, the number of user pairs $N$, and the number of reflecting elements $K$ on the secrecy performance. To circumvent this difficulty, we are motivated to investigate the asymptotic performance analysis based on extreme theory \cite{Shahab_TWC2007}, as shown in the following corollary.

\begin{corollary}
  For the proposed secure IRS assisted two-way communication scheme, we obtain: 1) $\bar{R}_{s_i}\varpropto\log P$, when $N$ and $K$ are finite values but $P\rightarrow\infty$. 2) $\bar{R}_{s_i}\varpropto2\log K$, when $N$ and $P$ are finite values but $K\rightarrow\infty$. 3) $\bar{R}_{s_i}\varpropto\log\log N$, when $K$ and $P$ are finite values but $N\rightarrow\infty$.
\end{corollary}
\begin{IEEEproof}
  In the large $P$ regime, it is readily verified that $\gamma_{n^\ast1}$ and $\gamma_{n^\ast2}$ increase with an increase in $P$, while $\gamma_{e1}$ and $\gamma_{e1}$ approach a constant (since $\gamma_{e1}<\gamma_{n^\ast1}$). Thus, an ASR scaling law with $P$ of $\log P$ is achieved for $s_1$ and $s_2$.

  Since $|h_{nk}|$ and $|g_{nk}|$ are independent Rayleigh variables with mean equal to $\frac{\pi}{2}$, and thus, we have $\mathbb{E}[|h_{nk}||g_{nk}|]=\frac{\pi}{4}$.  According to the CLT, we have $\sum_{k=1}^K|h_{nk}||g_{nk}|/K\rightarrow\frac{\pi}{4}$ for sufficiently large $K$. Using this result, we obtain that $\zeta_n\rightarrow\frac{\pi^2K^2}{16}$ as $K\rightarrow\infty$. By using the fact that $\max_{n=1,\dots,N}\zeta_n\geq\zeta_l$, $\forall l\in\{1,\dots,N\}$ and applying \cite[Theorem~4]{Shahab_TWC2007}, it is shown that $Q_{\text{M},1}$ and $Q_{\text{M},2}$ scale as $2\log K$. On the other hand, for sufficiently large $K$, $\mathbf{h}_e^T\Theta\mathbf{h}_n$ and $\mathbf{h}_e^T\Theta\mathbf{g}_n$ are approximated as complex Gaussian random variables with zero mean and variance of $K$ \cite{ZDing_WCL2020}. Hence, we have $\mathbb{E}[|\varphi_{ne}|^2],\mathbb{E}[|\psi_{ne}|^2]\rightarrow K$ and $\mathbb{E}[\gamma_{e1}],\mathbb{E}[\gamma_{e2}]\rightarrow c$ as $K\rightarrow\infty$, where $c$ is a constant. Consequently, an ASR scaling law with $K$ of $2\log K$ is achieved for $s_1$ and $s_2$.

  In the large $N$ regime, according to extreme theory \cite{Shahab_TWC2007}, it can be shown that $\gamma_{n^\ast1}$ and $\gamma_{n^\ast2}$ behave like $P\log N+O(\log\log N)$. Applying \cite[Theorem~4]{Shahab_TWC2007}, we obtain that $Q_{\text{M},1}$ and $Q_{\text{M},2}$ scale as $\log\log N$. Moreover, both $\gamma_{e1}$ and $\gamma_{e1}$ are independent of $N$ when $N$ is large. Thus, an ASR scaling law with $N$ of $\log\log N$ is achieved for $s_1$ and $s_2$.
\end{IEEEproof}

\begin{figure*}[t]
  \normalsize
  \centering
  \begin{minipage}[t]{0.32\textwidth}
    \centering
    \includegraphics[width=2.5in]{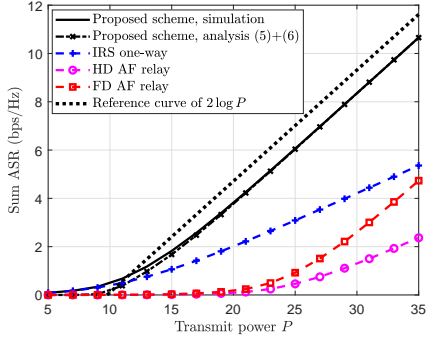}
    \caption{Impact of $P$ on the secrecy performance.}
    \label{simfig1}
  \end{minipage}
  \begin{minipage}[t]{0.32\textwidth}
    \centering
    \includegraphics[width=2.5in]{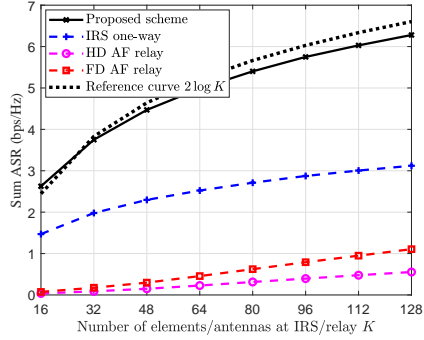}
    \caption{Impact of $K$ on the secrecy performance.}
    \label{simfig2}
  \end{minipage}
  \begin{minipage}[t]{0.32\textwidth}
    \centering
    \includegraphics[width=2.5in]{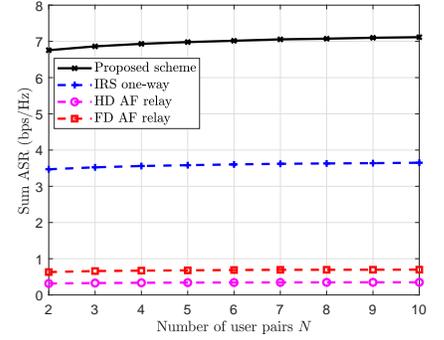}
    \caption{Impact of $N$ on the secrecy performance.}
    \label{simfig3}
  \end{minipage}
\end{figure*}

Considering non-secrecy IRS assisted two-way communications, similar scaling laws can be obtained. This verifies the efficiency of the proposed scheme for security guarantee.

Corollary 1 reveals an important design insight: Increasing $P$, $N$, and $K$ are helpful to the secrecy performance of IRS assisted two-way communications, but the impact of $N$ is less than $P$ and $K$ since the growth with $N$ is $\log\log(\cdot)$ while it is $\log(\cdot)$ for $P$ and $2\log(\cdot)$ for $K$. For instance, larger $K$ helps to achieve higher beamforming gain at the end users as well as strong jamming strength at Eve, and larger $N$ provides higher multiuser diversity gain.

\section{Numerical Results}
\label{sec:simulation}

We assume that the IRS is located at $(15,0)$ meter (m), Eve is located at $(15,20)$ m, $\text{A}_n$ and $\text{B}_n$ are uniformly deployed inside a disc centered at $(0,0)$~m and $(30,0)$ m, respectively, with radius equal to $5$ m. Moreover, $G_u=G_e=15$ dBi, $S=0.1\ \text{m}^2$ $\sigma_0^2=-70$~dBm, $\sigma_l^2=-40$~dBm, $\alpha=3$, and $M=20$. The following baseline schemes are considered for comparison: (1) IRS one-way with user-aided jamming, where the information exchange is completed in two phases, and in each phase one user transmits its data signal and the other user transmits a jamming signal. (2) Full-duplex (FD) or half-duplex (HD) amplify-and-forward (AF) relay aided two-way communications. The sum ASR of FD is $\bar{R}_\text{FD}=\sum_{i=1}^2\{\mathbb{E}[\log(1+\gamma_{n,i})] -\mathbb{E}[\log(1+\gamma_{e,i})]\}^+$, where $\gamma_{n,1}=\frac{\beta_\text{AR}\beta_\text{BR}\kappa_\text{f}^2P \|\mathbf{h}_n\|^2\|\mathbf{g}_n\|^2}{(\beta_\text{AR}\kappa_\text{f}^2\|\mathbf{h}_n\|^2+1) (\sigma_l^2+\sigma_0^2)}$, $\gamma_{n,2}=\frac{\beta_\text{AR}\beta_\text{BR}\kappa_\text{f}^2P \|\mathbf{h}_n\|^2|\mathbf{g}_n\mathbf{u}|^2}{(\beta_\text{BR}\kappa_\text{f}^2 |\mathbf{g}_n\mathbf{u}|^2+1)(\sigma_l^2+\sigma_0^2)}$, $\gamma_{e,1}=\frac{P(\sqrt{\beta_\text{Ae}}h_{ne}+\sqrt{\beta_\text{AR}\beta_\text{Re}} \kappa_\text{f}\|\mathbf{h}_n\||\mathbf{h}_e\mathbf{u}|)^2} {P(\sqrt{\beta_\text{Be}}g_{ne}+\sqrt{\beta_\text{BR}\beta_\text{e}} \kappa_\text{f}\|\mathbf{g}_n\||\mathbf{h}_e\mathbf{u}|)^2 +\beta_\text{e}\kappa_\text{f}^2|\mathbf{h}_e\mathbf{u}|^2 (\sigma_l^2+\sigma_0^2)+\sigma_0^2}$, $\gamma_{e,2}=\frac{P(\sqrt{\beta_\text{Be}}g_{ne}+\sqrt{\beta_\text{BR}\beta_\text{e}} \kappa_\text{f}\|\mathbf{g}_n\||\mathbf{h}_e\mathbf{u}|)^2} {\beta_\text{e}\kappa_\text{f}^2|\mathbf{h}_e\mathbf{u}|^2 (\sigma_l^2+\sigma_0^2)+\sigma_0^2}$, $\mathbf{u}=\frac{\mathbf{h}_n^H}{\|\mathbf{h}_n\|}$, and $\kappa_\text{f}^2=\frac{P}{P\beta_\text{AR}\|\mathbf{h}_n\|^2 +P\beta_\text{BR}\|\mathbf{g}_n\|^2+\sigma_l^2+\sigma_0^2}$. Moreover, the sum ASR of HD is $\bar{R}_\text{HD}=\sum_{i=1}^2\frac12\{\mathbb{E}[\log(1+\tilde{\gamma}_{n,i})] -\mathbb{E}[\log(1+\tilde{\gamma}_{e,i})]\}^+$, where $\tilde{\gamma}_{n,1}= \frac{\rho_0\beta_\text{AR}\beta_\text{BR}\|\mathbf{h}_n\|^2 \|\mathbf{g}_n\|^2} {\beta_\text{AR}\|\mathbf{h}_n\|^2+\kappa_\text{h}^{-2}}$, $\tilde{\gamma}_{n,2}=\frac{\rho_0\beta_\text{AR}\beta_\text{BR} \|\mathbf{h}_n\|^2|\mathbf{g}_n\mathbf{u}|^2} {\beta_\text{BR}|\mathbf{g}_n\mathbf{u}|^2+\kappa_\text{h}^{-2}}$, $\tilde{\gamma}_{e,1}=\frac{\rho_0\beta_\text{Ae}|h_{ne}|^2} {\rho_0\beta_\text{Be}|g_{ne}|^2+1}+\frac{\rho_0\beta_\text{e}\beta_\text{AR} \|\mathbf{h}_n\|^2|\mathbf{h}_e\mathbf{u}|^2}{\rho_0\beta_\text{e}\beta_\text{BR} \|\mathbf{g}_n\|^2|\mathbf{h}_e\mathbf{u}|^2+\beta_\text{e} |\mathbf{h}_e\mathbf{u}|^2+\kappa_\text{h}^{-2}}$, $\tilde{\gamma}_{e,2}=\rho_0\beta_\text{Be}|g_{ne}|^2 +\frac{\rho_0\beta_\text{e}\beta_\text{BR}\|\mathbf{g}_n\|^2|\mathbf{h}_e\mathbf{u}|^2}
{\beta_\text{e}|\mathbf{h}_e\mathbf{u}|^2+\kappa_\text{h}^{-2}}$, and $\kappa_\text{h}=\kappa_\text{f}(\sigma_l^2=0)$.

Fig.~\ref{simfig1} shows the impact of transmit power $P$ on the secrecy performance with $K=32$ and $N=10$. The sum ASR is defined as $\bar{R}_{s_1}+\bar{R}_{s_2}$. A general trend from Fig.~\ref{simfig1} is that the sum ASRs of all the schemes increase as $P$ increases, and the proposed scheme achieves the largest sum ASR. The IRS assisted one-way communications with user-aided jamming scheme achieves a worse secrecy performance, due to the fact that it needs two phases to finish the bidirectional information exchange, which is spectrally inefficient. Interestingly, both FD and HD relay assisted two-way communications achieve the worst sum ASR. The reasons are explained as follows: 1) The multi-antenna relay suffers a smaller beamforming gain than that of the IRS (i.e., $\mathcal{O}(K)$ versus $\mathcal{O}(K^2)$)\cite{QingqingWu_CM2020}, thus achieving a lower legitimate rate. 2) Eve can employ maximal-ratio combining to decode the received signals, which may yield a higher eavesdropping rate. It is also observed that the ASR lower bound is close to the simulated ones in all $P$ regime, thus confirming the derived analytical results.

Figs.~\ref{simfig2} and \ref{simfig3} show the sum ASR as a function of the number of reflecting elements/antennas $K$ and the number of user pairs $N$. In Fig.~\ref{simfig2}, we assume $P=20$ dBm and $N=6$, and in Fig.~\ref{simfig3}, we assume $P=30$ dBm and $K=32$. It is observed from Fig.~\ref{simfig2} that the sum ASR is an increasing function of $K$ for all the schemes, due to the fact that more reflecting elements in the IRS provide a stronger cascaded channel for legitimate reception quality improvement. In particular, the proposed scheme yields the best secrecy performance. Moreover, its sum ASR has the same increasing slope as the reference curve, which indicates that a sum ASR scaling law of $4\log K$ is achieved, thus verifying the results in Corollary~1. Similar observations can be also seen from Fig.~\ref{simfig3}.

\section{Conclusion}

This letter proposed a secure IRS assisted multiuser two-way communication scheme. Analytical expressions of a lower bound on the ASR and its scaling laws were derived in closed form to evaluate the secrecy performance and gain valuable insights. Both analytical and simulated results showed that the proposed scheme achieve perfect security and its ASR of each signal increases at $\log P$, $2\log K$, and $\log\log N$ rates.

\section*{Appendix: Proof of Theorem 1}

By applying Lemma 1, we can easily obtain the CDF of $\zeta_{n^\ast1}$ as $F_{\zeta_{n^\ast1}}(x)=\frac{1}{\Gamma^N(K\mu)} \gamma^N\big(K\mu,\frac{\sqrt{x}}{\nu}\big)$. Using this result and the G-C quadrature, $Q_{\text{M},1}$ can be approximated by
\begin{align}
\label{AP2}
  Q_{\text{M},1}&=\sum_{n=1}^N\Pr\big(n^\ast=n\big)\rho_\text{AB}
  \int_0^\infty\frac{1-F_{\gamma_{n^\ast1}}(x)}{1+\rho_\text{AB}x}dx\nonumber\\
  &\approx\sum_{n=1}^N\widetilde{\sum\limits_{m,M}}\frac{\rho_\text{AB}\sec^2x_m (1-F_{\zeta_{n^\ast1}}(\tan x_m))} {N(1+\rho_\text{AB}\tan x_m)},
\end{align}
where we use the change of variable $x=\tan y$.

\begin{figure*}[t]
  \begin{align}
  \label{AP9}
    Q_{\text{E},2}&=\int_0^\infty\!\!\!\int_0^\infty\Bigg( \int_{\frac{y}{x+\frac1{\rho_0}}}^\infty\log\bigg(1+\frac{x}{y+\frac1{\rho_0}}\bigg) dF_{\gamma_{n^\ast1}}(z)+\!\int_0^{\frac{y}{x+\frac1{\rho_0}}}\log\big(1+\rho_0x\big) dF_{\gamma_{n^\ast1}}(z)\Bigg)f_{|\psi_{n^\ast e}|^2}(x) f_{|\varphi_{n^\ast e}|^2}(y)dxdy\nonumber\\
    &=\underbrace{\int_0^\infty\!\!\!\int_0^\infty \log\bigg(1+\frac{x}{y+\frac1{\rho_0}}\bigg) \frac{e^{-\frac{x}{\sigma_{e'}^2}-\frac{y}{\sigma_e^2}}}{\sigma_e^2\sigma_{e'}^2} dxdy}_{J_1}+\!\underbrace{\int_0^\infty\!\!\!\int_0^\infty\log\bigg(\frac{1+\rho_0y} {1+\frac{y}{x+\frac1{\rho_0}}}\bigg)F_{\gamma_{n^\ast1}} \bigg(\frac{y}{x+\frac1{\rho_0}}\bigg) \frac{e^{-\frac{x}{\sigma_{e'}^2}-\frac{y}{\sigma_e^2}}}{\sigma_e^2\sigma_{e'}^2} dxdy}_{J_2}.
  \end{align}
  \hrule\vspace{-2mm}
\end{figure*}

Furthermore, before deriving $Q_{\text{E},1}$, we need to first determine the CDFs of $|\varphi_{n^\ast e}|^2$ and $|\psi_{n^\ast e}|^2$. However, it is in general difficult to derive closed-form expressions for $\mathbf{h}_e^T\Theta\mathbf{h}_n$ and $\mathbf{h}_e^T\Theta\mathbf{g}_n$, since they are sums of complex-valued random variables, for which the real and imaginary parts are correlated. Thanks to \cite[Lemma~2]{ZDing_WCL2020}, for a large $K$, we can approximate them as complex Gaussian random variables with zero mean and variance of $K$. Based on this result, we can obtain that $\varphi_{ne}$ and $\psi_{ne}$ follow a complex Gaussian distribution with zero mean and variances of $\sigma_e^2=\beta_\text{I-Ae}K+\beta_\text{D-Ae}$ and $\sigma_{e'}^2=\beta_\text{I-Be}K+\beta_\text{D-Be}$, respectively. Thus, the CDF of $|\varphi_{n^\ast e}|^2$ can be derived as $F_{|\varphi_{n^\ast e}|^2}(x)=\sum_{n=1}^N\Pr\big(n^\ast=n\big)\big(1-e^{-\frac{x}{\sigma_e^2}}\big) =1-e^{-\frac{x}{\sigma_e^2}}$. Similarly, the CDF of $|\psi_{n^\ast e}|^2$ is given by $F_{|\psi_{n^\ast e}|^2}(x)=1-e^{-\frac{x}{\sigma_{e'}^2}}$. Based on the above, the CDF of $\gamma_{e1}$ is obtained as $F_{\gamma_{e1}}(x)=1-\frac{\sigma_e^2}{\sigma_e^2+\sigma_{e'}^2x} e^{-\frac{x}{\rho_0\sigma_e^2}}$. Using this result and \cite[eq.~(3.352.4)]{Table}, $Q_{\text{E},1}$ is derived as
\begin{align}
\label{AP8}
  Q_{\text{E},1}&=\frac{\sigma_e^2}{\sigma_{e'}^2}\int_0^\infty \frac{e^{-\frac{x}{\rho_0\sigma_e^2}}}{(1+x)(\sigma_0^2/\sigma_{e'}^2+x)}dx\nonumber\\
  &=\frac{\sigma_e^2}{\sigma_e^2-\sigma_{e'}^2}\bigg(e^{\frac{1}{\rho_0\sigma_{e'}^2}} \mathrm{Ei}\Big(\frac{-1}{\rho_0\sigma_{e'}^2}\Big)
  -e^{\frac{1}{\rho_0\sigma_e^2}} \mathrm{Ei}\Big(\frac{-1}{\rho_0\sigma_e^2}\Big)\bigg).
\end{align}
Combining \eqref{AP2} with \eqref{AP8}, a closed-form $\bar{R}_{s_1}$ is obtained.

On the other hand, since $\gamma_{n^\ast1}$ and $\gamma_{n^\ast2}$ are symmetric, the CDF of $\gamma_{n^\ast2}$ is the same with $\gamma_{n^\ast1}$, and thus, $Q_{\text{M},2}$ can be derived in a similar manner with \eqref{AP2}. In addition, we can compute $Q_{\text{E},2}$ using \eqref{AP9}, shown at the top of the page, where $J_1$ is approximated by
\begin{align}
\label{AP10}
  J_1&\approx-\frac{\pi^2e^{-\frac1{\rho_0\sigma_{e'}^2}}}{4M\sigma_e^2} \widetilde{\sum\limits_{m,M}}e^{-\frac{\tan x_m}{\sigma_{e'}^2}-\frac{\tan x_m}{\sigma_e^2}} \nonumber\\
  &\quad\times\mathrm{Ei}\Big(-\frac{\tan x_m+1/\rho_0}{\sigma_e^2}\Big)\sec^2x_m.
\end{align}
To calculate $J_2$, we first define $\alpha=\frac{y}{x+\frac1{\rho_0}}$, $\beta=\rho_0y$, and construct a Jacobian matrix as follows
\begin{equation}
\label{AP11}
  \mathbf{P}=\bigg[\begin{array}{cc}
    \frac{dx}{d\alpha} & \frac{dx}{d\beta}\\
    \frac{dy}{d\alpha} & \frac{dy}{d\beta}
  \end{array}\bigg]=\bigg[\begin{array}{cc}
    -\frac{\beta}{\rho_0\alpha^2} & \frac{1}{\rho_0\alpha}\\
    0 & \frac1{\rho_0}
  \end{array}\bigg],
\end{equation}
where $|\mathbf{P}|=-\frac{\beta}{\rho_0^2\alpha^2}$. Then, $J_2$ can be approximated using \cite[eq.~(3.351.3)~and~(4.337.5)]{Table} and the G-C quadrature by
\begin{align}
\label{AP12}
  J_2&=\frac{e^{-\frac1{\rho_0\sigma_{e'}^2}}}{\rho_0^2\sigma_e^2\sigma_{e'}^2} \int_0^\infty\!\!\!\int_0^\infty\log\Big(\frac{1+\beta}{1+\alpha}\Big) \frac{\beta F_{\gamma_{n^\ast1}}(\alpha)}{\alpha^2}e^{-\tau(\alpha)\beta}d\alpha d\beta\nonumber\\
  &\approx\frac{\pi^2e^{-\frac1{\rho_0\sigma_{e'}^2}}} {4M\rho_0^2\sigma_e^2\sigma_{e'}^2}\widetilde{\sum\limits_{m,M}} \frac{\Delta(\tan x_m)F_{\gamma_{n^\ast1}}(\tan x_m)}{\tau^2(\tan x_m)\csc^{-2}x_m},
\end{align}
where $\Delta(\alpha)=\big(\frac1{\tau(\alpha)}-1\big)e^{\tau(\alpha)} \mathrm{Ei}(-\tau(\alpha))-\log(1+\alpha)+1$ and $\tau(\alpha)=\frac1{\rho_0\sigma_{e'}^2\alpha}+\frac1{\rho_0\sigma_e^2}$. Combining \eqref{AP2}, \eqref{AP10}, and \eqref{AP12}, a closed-form expression for $\bar{R}_{s_2}$ is derived, which completes the proof of Theorem 1.

\end{document}